\documentclass[prd,aps,showpacs,showkeys,preprint,floats,eqsecnum,amsmath]{revtex4}

\begin{document}

\title{Electrodynamics of moving magnetoelectric media: variational approach}

\author{Yuri N.~Obukhov}
\email{yo@htp.uni-koeln.de}
\affiliation{Institute for Theoretical Physics, University of Cologne,
50923 K\"oln, Germany}
\affiliation{Department of Theoretical Physics,
Moscow State University, 117234 Moscow, Russia}

\author{Friedrich W. Hehl}
\email{hehl@thp.uni-koeln.de}
\affiliation{Institute for Theoretical Physics, University of Cologne,
50923 K\"oln, Germany}
\affiliation{Dept.\ of Phys.\ Astron., 
University of Missouri-Columbia, Columbia, MO 65211, USA}

\begin{abstract}
Recently, Feigel has predicted a new effect in magnetoelectric media. 
The theoretical evaluation of this effect requires a careful analysis
of a dynamics of the moving magnetoelectric medium and, in particular,
the derivation of the energy-momentum of the electromagnetic field in
such a medium. Then, one can proceed with the study of the wave propagation 
in this medium and derive the mechanical quantities such as the energy,
the momentum, and their fluxes and the corresponding forces. 
In this paper, we develop a consistent general-relativistic
variational approach to the moving dielectric and magnetic medium with 
and without magnetoelectric properties. The old experiments in which the 
light pressure was measured in fluids are reanalysed in our new framework. 
\end{abstract}
\maketitle

\section{Introduction}

The discussion of the electrodynamics of moving media has a long history. 
At present, the general structure of classical electrodynamics appears to
be well established. In particular, in the generally covariant pre-metric 
approach to electrodynamics \cite{Schouten,Post,HO02,Lindell,Delph,Toupin}, the
electric charge and the magnetic flux conservation laws manifest themselves
in the Maxwell equations for the excitation $H=({\cal D},{\cal H})$
and the field strength $F=(E,B)$, namely $dH = J,\ dF = 0$. These
equations should be supplemented by a constitutive law $H = H(F)$. The
latter relation contains the crucial information about the underlying
physical continuum (i.e., spacetime and/or material medium), in particular,
about the spacetime metric.
Mathematically, this constitutive law arises either from a suitable
phenomenological theory of a medium or from the electromagnetic field
Lagrangian. It can be a nonlinear or even nonlocal relation between
the electromagnetic excitation and the field strength. The
constitutive law is called a spacetime relation if it applies to
spacetime (``the vacuum'') itself.

Among many physical applications of classical electrodynamics, the
problem of the interaction of the electromagnetic field with matter
occupies a central position. The fundamental question, which arises
in this context, is about the definition of the energy and momentum in
the possibly moving medium.  The discussion of the energy-momentum
tensor in macroscopic electrodynamics is quite old. The beginning of
this dispute dates back to Minkowski \cite{minkowski}, Einstein and Laub 
\cite{laub}, and Abraham \cite{abraham}.
Nevertheless, up to now the question was not settled and there is an on-going 
exchange of conflicting opinions concerning the validity of the Minkowski 
versus the Abraham energy-momentum tensor, see, e.g., the review 
\cite{Brevik}. Even experiments were not quite 
able to make a definite and decisive choice of electromagnetic energy and 
momentum in material media. A consistent solution of this problem has been
proposed in \cite{emt,HO02} (cf. also the earlier work \cite{poincelot}) in 
the context of a new axiomatic approach to electrodynamics.

Recently Feigel \cite{feigel} has studied, theoretically, the dynamics of 
a dielectric magnetoelectric medium in an external electromagnetic field.
He predicted that the contributions of the quantum vacuum waves (or 
``virtual photons") could transfer a nontrivial momentum to matter. This
prediction was made with the help of the non-relativistic formalism. 
In our opinion, a proper relativistic analysis is needed for a better
understanding of the physics and of the viability of this phenomenon. Here
we begin to reconsider this problem in a covariant framework as developed 
earlier in \cite{HO02,emt}. As a first step, we develop a variational 
approach to the description of the dynamics of a moving magnetoelectric 
medium. The corresponding energy-momentum of matter plus electromagnetic
field that arises can be derived straightforwardly in this formalism from
the variation of the total action with respect to the spacetime metric.

\section{Preliminaries: the essence of the Feigel effect}\label{SecIntro}

The Feigel effect \cite{feigel} can be described in simple terms as follows:
Let us consider an isotropic homogeneous medium with the electric and 
magnetic constants $\varepsilon, \mu$. Electromagnetic waves are propagating
in such a medium absolutely symmetrically, with the Fresnel equation 
describing the unique light cone. This is easily derived from the 
constitutive relations ${\cal D} = \varepsilon\varepsilon_0\,E$ and 
${\cal H} = (\mu\mu_0)^{-1}B$. 

However, if a medium is placed in crossed constant external electric
and magnetic fields, then it acquires magnetoelectric properties. As
a result, we have the {\it anisotropic} magnetoelectric medium with 
$\varepsilon, \mu$, plus the magnetoelectric matrix $\beta$ (determined 
by the external fields) which modifies the constitutive relations
to ${\cal D} = \varepsilon\varepsilon_0\,E + \beta\cdot B$ and ${\cal H} = 
(\mu\mu_0)^{-1}B - {\beta}^{\rm T}\cdot E$; here $^{\rm T}$ denotes the 
transposed matrix.

Accordingly, the wave propagation in such a medium also becomes an\-isotropic
and bi\-refringent, with the wave covectors now belonging to two light 
cones. Applying this to vacuum waves (or, perhaps, better to say to the 
``vacuum fluctuations" or ``virtual photons") propagating in the magnetoelectric 
body, Feigel \cite{feigel} computed the total momentum carried by these waves 
and concluded that it is nontrivial. In accordance with this derivation, a 
body should move with a small but non-negligeable velocity. Earlier the 
Feigel process was discussed in \cite{comm1,reply1,comm2,reply2,Tiggelen}.

In order to evaluate the possible Feigel effect, it is necessary 
to substitute the ``vacuum waves" into the energy-momentum tensor. This 
paper is devoted to the derivation of the latter in the framework of a
variational approach.

\section{Constitutive relation}\label{ConstRel}

Within the axiomatics of the premetric generally covariant framework \cite{HO02},
the projection technique is used to define the electric and magnetic phenomena 
in an arbitrarily {\it moving} medium. As in \cite{HO02}, we assume that the 
spacetime is foliated into spatial slices with time $\sigma$ and 
transverse vector field $n$. 

When applying the projection technique to the 2-forms of the electromagnetic
excitation $H$ and and the electromagnetic field strength $F$, we obtain the 
three-dimensional objects: the magnetic and electric excitations ${\cal H}$ 
and ${\cal D}$ as longitudinal and transversal parts of $H$ and, similarly, 
electric and magnetic fields $E$ and $B$ as longitudinal and transversal parts 
of $F$, respectively, namely 
\begin{equation}
H = -{\cal H}\wedge d\sigma + {\cal D}\qquad {\rm and}\qquad 
F = E\wedge d\sigma + B.
\end{equation}
This foliation is called the {\it laboratory} foliation, with the coordinate 
time variable $\sigma$ labeling the slices of this foliation. 

The spacetime metric ${\mathbf g}$ introduces the scalar product in the 
tangent space and defines the line element. With respect 
to the laboratory foliation coframe it reads ($a,b,... = 1,2,3$)
\begin{equation}
ds^2 = N^2\,d\sigma^2 + g_{ab}\,\underline{dx}^a\,\underline{dx}^b 
= N^2\,d\sigma^2 - {}^{\hbox{$\scriptstyle{(3)}$}}g_{ab}
\,\underline{dx}^a\,\underline{dx}^b.\label{metF}
\end{equation} 
Here $N^2 = {\mathbf g}(n,n)$ is the length square of the foliation vector
field $n$, and $\underline{dx}^a = dx^a - n^a\,d\sigma$ is the transversal
3-covector basis, in accordance with the definitions above. The 3-metric
${}^{\hbox{$\scriptstyle{(3)}$}}g_{ab}$ is the positive definite Riemannian
metric on the spatial 3-dimensional slices corresponding to fixed values 
of the time $\sigma$. This metric defines the 3-dimensional Hodge duality
operator ${}^{\underline{\star}}$. 

The constitutive relation which links the electromagnetic field strength to
the electromagnetic excitation, $H = H(F)$, can be nonlocal and nonlinear,
in general. Here we will confine our attention to the local and linear  
constitutive relation. 

Then, if we write the the excitation 2-form in terms of its components in
a local coordinate system $\{x^i\}$, $({\cal H}, {\cal D}) = H=H_{ij}
\,dx^i\wedge dx^j/2$ (with $i,j,\dots = 0,1,2,3$), the local and linear 
constitutive relation means that the components of the excitation are
local linear functions of the components of the field strength $(E, B) = 
F=F_{ij}\,dx^i \wedge dx^j/2$:
\begin{equation}\label{HchiF}
H=\kappa(F)\,,\qquad H_{ij}={\frac 1 2}\,\kappa_{ij}{}^{kl}\,F_{kl}\,.
\end{equation}
Along with the original constitutive $\kappa$-tensor, it is convenient to 
introduce an alternative representation of the constitutive tensor:
\begin{equation}
\chi^{ijkl} := {\frac 1 2}\,\epsilon^{ijmn}\,\kappa_{mn}{}^{kl}.\label{chikap}
\end{equation}

Performing a $(1+3)$-decomposition of covariant electrodynamics, as 
described above, we can write $H$ and $F$ as column 6-vectors with the
components built from the magnetic and electric excitation 3-vectors
${\cal H}_a, {\cal D}^a$ 
and the electric and magnetic field strengths $E_a, B^a$, respectively. 
Then the linear spacetime relation (\ref{HchiF}) reads:
\begin{equation}
  \left(\begin{array}{c} {\cal H}_a \\ {\cal D}^a\end{array}\right) 
= \left(\begin{array}{cc} {{\cal C}}^{b}{}_a & {{\cal B}}_{ba} \\ 
{{\cal A}}^{ba}& {{\cal D}}_{b}{}^a \end{array}\right) \left(
\begin{array}{c} -E_b\\  {B}^b\end{array}\right)\,.\label{CR'}
\end{equation}
Here the constitutive tensor is conveniently represented by the 
$6\times6$-matrix
\begin{equation}\label{kappachi}
\kappa_I{}^K=\left(\begin{array}{cc} {{\cal C}}^{b}{}_a & {{\cal B}}_{ba} \\ 
{{\cal A}}^{ba}& {{\cal D}}_{b}{}^a \end{array}\right)\,,\qquad 
\chi^{IK}= \left( \begin{array}{cc} {\cal B}_{ab}& {\cal D}_a{}^b \\ 
{\cal C}^a{}_b & {\cal A}^{ab} \end{array}\right)\,.
\end{equation}
Assuming that the {\it skewon} and the {\it axion} pieces are {\it absent}, 
we find that the constitutive matrices satisfy ${\cal A}^{ab} = {\cal A}^{ba}$, 
${\cal B}_{ab} = {\cal B}_{ba}$, and ${\cal D}_b^{\ a} = {\cal C}^a{}_b$,
with ${\cal C}^a{}_a = 0$.

The dynamics of a material medium is encoded in the structure of another 
foliation $(\tau, u)$ which is determined by the four-vector field of the 
velocity $u$ of matter and the proper time coordinate $\tau$. Accordingly, 
we have to formulate the constitutive law with respect to this, so called 
{\it material foliation}. As a first step, we observe that the relation between 
the two coframe bases, namely those of the laboratory foliation $(d\sigma, 
\underline{dx}^a)$ and of the material foliation $(d\tau, 
{\underset{\widetilde{\ \ \ }}{dx}}^a)$ is as follows \cite{HO02}:
\begin{equation}
\left(\begin{array}{c}d\sigma \\ \underline{dx}^a\end{array}\right) =
\left(\begin{array}{c|c}\gamma c/N & v_b/(cN)\\ \hline\gamma v^a & 
\delta^a_b\end{array}\right)\left(\begin{array}{c}d\tau\\ {\underset  
{\widetilde{\ \ \ }}  {dx}}^b\end{array}\right).\label{dsigmaM}
\end{equation}
Here, for the {\it relative velocity} 3-vector, we introduced the notation
\begin{equation}
v^a:={\frac c N}\left({\frac {u^a}{u^{(\sigma)}}}-n^a\right),\quad {\rm with}
\quad\gamma:={\frac 1 {\sqrt{1 - {\frac {v^2} {c^2}}}}}\,.\label{gamM}
\end{equation}
Substituting (\ref{dsigmaM}) into (\ref{metF}), we find for the line
element in terms of the new variables
\begin{equation}
ds^2 = c^2\,d\tau^2 - {\widehat g}_{ab}\,{\underset {\widetilde{\ \  \ }} 
{dx}}^a\,{\underset {\widetilde{\ \ \ }}{dx}}^b,\quad{\rm where}\quad 
\widehat{g}_{ab} = {}^{(3)}g_{ab} - {\frac 1 {c^2}}\,v_av_b.\label{metM}
\end{equation}
The metric $\widehat{g}_{ab}$ of the material foliation has the inverse
\begin{equation}\label{invghat}
\widehat{g}^{ab} = {}^{(3)}g^{ab} + {\frac {\gamma^2} {c^2}}\,v^av^b,
\end{equation}
where ${}^{(3)}g^{ab}$ is the inverse of ${}^{(3)}g_{ab}$. For the determinant 
one finds $(\det\widehat{g}_{ab}) = (\det {}^{(3)}g_{ab})\,\gamma^{-2}$. We will 
denote the 3-dimensional Hodge star defined by the metric 
$\widehat{g}_{ab}$ as ${}^{\widehat{\ast}}$. 

Given the transformation (\ref{dsigmaM}) between the 1-form bases of the 
two foliations (laboratory and material), it is straightforward to calculate
the components of the constitutive tensor with respect to the moving 
reference frame from the constitutive tensor which describes the matter at 
rest. The corresponding explicit general transformation formulas for the 
3-dimensional matrices ${\cal A}, {\cal B}, {\cal C}$ can be found in
\cite{HO02}. A nontrivial outcome of this formalism is that the isotropic
moving medium is described by the constitutive relation based on the 
so-called {\it optical metric} (first introduced by Gordon \cite{gordon}).
We will use this fact in the subsequent derivations.

\section{Relativistic fluid}\label{Fluid}

The earlier work includes that for the ideal fluids \cite{taub,schutz,bailyn}
and the case of the fluids in electromagnetic fields was discussed in 
\cite{Penfield1,Penfield2,Mikura,tiersten}. 
Recently, a new study was reported in \cite{Dereli1,Dereli2}.
 
An ideal fluid which consists of structure-less elements (particles)
is characterized in the Eulerian approach by the  fluid 4-velocity $u^i$,
the internal energy density $\rho$, the particle density $\nu$, the entropy 
density $s$, and the identity (Lin) coordinate $X$. Normally, it is 
assumed that the motion of a fluid is such that the number of particles
is constant and that the entropy and the identity of the elements is
conserved. In other words,
\begin{eqnarray}
\nabla_i(\nu u^i) &=& 0,\label{numbercons}\\
u^i\partial_is &=& 0,\label{entrcons}\\
u^i\partial_iX &=& 0.\label{idcons}
\end{eqnarray}  
By the conservation of the entropy only the reversible processes are allowed.
In a variational approach to continuous media, these assumptions are 
considered as constraints imposed on the dynamics of the fluid by means of
Lagrange multipliers. Then the fluid Lagrangian \cite{taub,schutz} reads 
$V_{\rm mat} = L_{\rm mat}\,\sqrt{-g}\,d\sigma\wedge\widehat{\epsilon}$ with 
($\widehat{\epsilon}=$ spatial volume 3-form)
\begin{equation}
L_{\rm mat} = -\,{\frac {\rho(\nu, s)} c} + \Lambda_0(u^iu_i - c^2) + 
\Lambda_1\nabla_i(\nu u^i)+\Lambda_2\,u^i\partial_is + \Lambda_3\,u^i
\partial_iX.\label{Lfluid}
\end{equation}
The Lagrange multipliers $\Lambda_1, \Lambda_2, \Lambda_3$ impose the 
constraints (\ref{numbercons})-(\ref{idcons}) on the dynamics of the fluid, 
whereas $\Lambda_0$ provides the standard normalization condition for 
the 4-velocity
\begin{equation}
g_{ij}u^iu^j = c^2.\label{u2}
\end{equation}
For the description of the thermodynamical properties of the fluid, the 
usual thermodynamical law (``Gibbs relation") is used,
\begin{equation}
T\,ds = d(\rho/\nu) + p\,d(1/\nu),\label{thermod}
\end{equation}
where $T$ is the temperature and $p$ the pressure. 

When the medium possesses dielectric and magnetic properties, then we have to 
add the Lagrangian of the electromagnetic field 
\begin{eqnarray}
V_{\rm em} &=& -\,{\frac 12}\,H\wedge F =  -\,{\frac 12}\,d\sigma\wedge
\left({\cal H}\wedge B - {\cal D}\wedge E\right)\nonumber\\ 
&=&  -\,{\frac 12}\,d\sigma\wedge\widehat{\epsilon}\left({\cal A}^{ab}
E_aE_b + {\cal B}_{ab}B^aB^b - 2{\cal C}^a{}_bE_aB^b\right).\label{Vem}
\end{eqnarray}

At first, let us consider the moving isotropic dielectric fluid without 
magnetoelectric properties, that is, ${\cal C}^a{}_b =0$. Then the 
electromagnetic field Lagrangian reads $V_{\rm em} = L_{\rm em}\,\sqrt{-g}
\,d\sigma\wedge\widehat{\epsilon}$ with 
\begin{equation}
L_{\rm em} = -\,{\frac {\lambda_0}{4\mu}}\,g_{\rm opt}^{ij}\,g_{\rm opt}^{kl}
\,F_{ik}F_{jl}.\label{Lem}
\end{equation}
Here $\lambda_0$ is the vacuum admittance, and
\begin{equation}
g_{\rm opt}^{ij} = g^{ij} - {\frac {1 - n^2}{c^2}}\,u^i\,u^j
\end{equation}
is the optical metric of Gordon \cite{gordon}.

\subsection{Equations of motion without a magnetoelectric effect}

At first, we consider the case when the fluid does not have magnetoelectric
properties: ${\cal C}^a{}_b =0$. We also assume that there are no free 
charges and currents in the fluid. 

The dynamics of the system (fluid$+$field) is thus described by the 
total Lagrangian $V = V_{\rm mat} + V_{\rm em}$. The corresponding equations
of motion are then derived from the variational principle for $V$ with the
independent variables $u^i, \nu, s, X, \Lambda_0, \Lambda_1, \Lambda_2,
\Lambda_3$, and $A_i$ (the covector of the electromagnetic potential). 

Variation with respect to the Lagrange multipliers $\Lambda_0, \Lambda_1, 
\Lambda_2, \Lambda_3$ yields the constraints (\ref{numbercons})-(\ref{idcons})
and (\ref{u2}), whereas variation of $V$ with respect to $\nu, s, X$ yields,
respectively,
\begin{eqnarray}
\nu u^i\partial_i\Lambda_1 + {\frac {\rho + p} c} &=& 0,\label{dVdnu}\\
\nabla_i\left(\Lambda_2 u^i\right) + {\frac {T\nu} c} &=& 0,\label{dVds}\\
\nabla_i\left(\Lambda_3 u^i\right) &=& 0.\label{dVdX}
\end{eqnarray}
In order to derive these results, we used the thermodynamical law 
(\ref{thermod}), from which one has $\partial\rho/\partial s = \nu T$ and
$\partial\rho/\partial \nu = (\rho + p)/\nu$. 

Finally, the variation of $V$ with respect to the 4-velocity yields
\begin{equation}
2\Lambda_0u_i - \nu\partial_i\Lambda_1 + \Lambda_2\partial_is + 
\Lambda_3\partial_iX + {\frac {\lambda_0(1 - n^2)}{\mu c^2}}\,F_{ik}
\,F^{jk}u_j = 0.\label{dVdu}
\end{equation}
The last term emerges from $V_{\rm em}$. Contracting the last equation with
$u^i$ and making use of (\ref{numbercons})-(\ref{idcons}), (\ref{u2}),
and (\ref{dVdnu}), we find
\begin{equation}
2\Lambda_0 = -\,{\frac 1 {c^2}}\left({\frac {\rho + p} c} + 
{\frac {\lambda_0(1 - n^2)}{\mu c^2}}\,F_{ik}u^kF^{il}u_l\right).\label{Lam0}
\end{equation}
The variation with respect to $A_i$ gives the Maxwell equations for the 
moving fluid
\begin{equation}\label{MaxF}
\nabla_j\left(g_{\rm opt}^{jk}\,g_{\rm opt}^{il}\,F_{kl}\right) = 0.
\end{equation}
When the fluid has free currents and charges, we have to add the interaction 
term $A_iJ^i$ to the electromagnetic field Lagrangian $L_{\rm em}$, and then 
the right-hand side of (\ref{MaxF}) picks up a nontrivial current $J^i$.

\subsection{Energy-momentum tensor}

The above consideration is not only relativistically covariant, but actually 
{\it generally} covariant. The metric $g_{ij}$ above describes an arbitrary 
curved spacetime. The invariance of the total Lagrangian $L$ under general 
coordinate transformations (using the standard Noether machinery) yields the
conservation of the energy-momentum of the system (fluid$+$field). We
obtain the energy-momentum tensor as usual from the variation of the 
total Lagrangian with respect to the metric, i.e.,
\begin{equation}
T_{ij} := 2c\,{\frac {\delta\left[\sqrt{-g}(L_{\rm mat} + L_{\rm em})\right]}
{\sqrt{-g}\delta g^{ij}}}. 
\end{equation}
The computation is straightforward, and it yields for (\ref{Lfluid}) and
(\ref{Lem})
\begin{eqnarray}
T_i{}^j &=& -\,\delta_i^j\,p + {\frac {p + \rho}{c^2}}\,u_iu^j + 
{\frac {1 - n^2}{\mu_0\mu\,c^4}}\,u_iu^j\,F_{mk}u^kF^{ml}u_l\nonumber\\
&& + {\frac 1 {\mu_0\mu}}\Bigg[-\,F_{ik}\,F^{jk} + {\frac 1 4}\,\delta_i^j
F_{kl}F^{kl} \nonumber\\ 
&&  + {\frac {1 - n^2}{c^2}}\left( F_{ik}u^kF^{jl}u_l
- {\frac 12}\,F_{mk}u^kF^{ml}u_l\,\delta_i^j\right)\Bigg].\label{Tij}
\end{eqnarray}
In the course of its derivation, we made use of the results of the previous
subsection, in particular of the equations (\ref{dVdnu}) and (\ref{Lam0}).
The first line in (\ref{Tij}) originates from $V_{\rm mat}$, whereas the
rest comes from $V_{\rm em}$. It is interesting though that it is
possible to rearrange the contributions of the fluid and of the field 
in such a way that the total energy-momentum is recast into the nice form
\begin{eqnarray}
T_i{}^j &=& -\,\delta_i^j\,p_{\rm eff} + {\frac {p_{\rm eff} 
+ \rho_{\rm eff}}{c^2}}\,u_iu^j\nonumber\\
&& + {\frac 1 {\mu_0\mu}}\left[-\,F_{ik}\,F^{jk} + {\frac 1 4}\,\delta_i^j
\,F_{kl}F^{kl} + {\frac {1 - n^2}{c^2}}\,F_{ik}u^kF^{jl}u_l
\right].\label{Teff}
\end{eqnarray}
Here we introduced the effective pressure and the effective energy density
by means of
\begin{eqnarray}
p_{\rm eff} &:=& p + {\frac {1 - n^2}{2\mu\mu_0\,c^2}}
\,F_{ij}u^jF^{ik}u_k,\label{Peff}\\  \rho_{\rm eff}&:=&\rho + 
{\frac {1 - n^2}{2\mu\mu_0\,c^2}}\,F_{ij}u^jF^{ik}u_k.\label{Reff}
\end{eqnarray}

General covariance underlies the conservation law of the total 
energy-momentum tensor
\begin{equation}
\nabla_j\,T_i{}^j = 0,
\end{equation}
which yields the equations of motion of the medium.

\subsection{Extension to the magnetoelectric case}

The above results all refer to the isotropic dielectric and magnetic medium
in motion without magnetoelectric crossterms. The extension to the 
magnetoelectric case, see O'Dell \cite{Odell}, is straightforward. The 
electromagnetic field Lagrangian (\ref{Lem}) is replaced by the more general one
\begin{equation}
L_{\rm em} = -\,{\frac {\lambda_0}{4\mu}}\,g_{\rm opt}^{ij}\,g_{\rm opt}^{kl}
\,F_{ik}F_{jl} - {\frac 1{2c^2}}\,\beta^{i'}{}_{j'}\,F_{i'i}u^i
\,\eta^{j'klm}F_{lm}u_k.\label{genLem}
\end{equation}
Here the tensor $\beta^i{}_j$ gives the relativistic generalization of the
magnetoelectric matrix $\beta$. The totally antisymmetric Levi-Civita tensor
is defined as usual by $\eta^{ijkl} = \epsilon^{ijkl}/\sqrt{-g}$, with the 
permutation symbol $\epsilon^{ijkl}$ that have the only nontrivial 
component, $\epsilon^{0123} = 1$. The dimension $[\beta^i{}_j] = [\lambda_0]$.

The tensor $\beta^i{}_j$ is assumed to be {\it traceless}. If a nontrivial 
trace is included, this will describe an {\it axion} contribution. Indeed, 
the trace $\beta^i{}_j = \delta^i_j\,\widetilde{\alpha}$, when substituted into 
(\ref{genLem}), yields the term $\sim \widetilde{\alpha}F_{i'i}u^i\eta^{i'klm}
F_{lm}u_k$. Using the identity (which states that a totally antisymmetric 
tensor of the 5-th rank is zero in four dimensions) $2u^{[i}\eta^{i']klm} 
= u^k\eta^{i'ilm} + u^l\eta^{i'kim} + u^m\eta^{i'kli}$, we find 
$\widetilde{\alpha}\,F_{i'i}u^i\eta^{i'klm}F_{lm}u_k = {\frac {c^2}4}
\,\widetilde{\alpha}\,\eta^{ijkl}F_{ij}F_{kl}$. Thus, indeed the trace 
$\widetilde{\alpha}$ adds an axion contribution to the electromagnetic
Lagrangian.  We assume that the axion is absent in this paper.

Furthermore, it is easy to see that only a projection of the magnetoelectric 
matrix on the rest frame of the 4-velocity $u$ enters the Lagrangian 
(\ref{genLem}), namely
\begin{equation}
- {\frac 1{2c^2}}\,\beta^{i'}{}_{j'}\,F_{i'i}u^i\eta^{j'klm}F_{lm}u_k = 
- {\frac 1{2c^2}}\,\widetilde{\beta}^{i'}{}_{j'}\,F_{i'i}u^i\eta^{j'klm}F_{lm}u_k, 
\end{equation}
where 
\begin{equation}
\widetilde{\beta}^{i}{}_{j} = P^i_kP^l_j\,\beta^k{}_l,
\end{equation}
with the projector defined by $P^i_k := \delta^i_k - u^iu_k/c^2$. By construction,
$u^j\widetilde{\beta}^{i}{}_{j} = 0$ and $u_i\widetilde{\beta}^{i}{}_{j} = 0$,
which means that only the 3-dimensional transversal part of the magnetoelectric
tensor contributes. 

Computation of the electromagnetic excitation $H = -\,\partial V_{\rm em}/
\partial F$ is straightforward. In the (1+3)-decomposed form, this yields the
constitutive relation (\ref{CR'}), where the 3-dimensional matrices read
explicitly:
\begin{eqnarray}
{\cal A}^{ab} &=& -\,{\stackrel {o}{\varepsilon}}{}^{ab} -\,{\stackrel {\rm me}
{\varepsilon}}{}^{ab},\label{Atot}\\
{\cal B}_{ab} &=& ({\stackrel {o}{\mu}}{}^{-1})_{ab} + ({\stackrel {\rm me}{\mu}}
{}^{-1})_{ab},\\
{\cal C}^a{}_b &=& {\stackrel {o}{\gamma}}{}^a{}_b + {\stackrel {\rm me}{\gamma}}
{}^a{}_b. 
\end{eqnarray}
Here the first terms on the right-hand sides describe the contributions of 
the isotropic moving medium
\begin{eqnarray}
{\stackrel {o}{\varepsilon}}{}^{ab} &=& {\frac {\lambda\,\sqrt{{}^{(3)}g}}
{c(1 - v^2/c^2)}}\left[{}^{(3)}g^{ab}\left(n - {\frac {v^2}{nc^2}}\right)
+ {\frac {v^av^b}{c^2}}\left({\frac 1 n} - n\right)\right],\\
({\stackrel {o}{\mu}}{}^{-1})_{ab} &=& {\frac {\lambda\,c}{\sqrt{{}^{(3)}g}
(1 - v^2/c^2)}}\left[{}^{(3)}g_{ab}\left({\frac 1 n} - {\frac {nv^2}{c^2}}\right)
+ {\frac {v^av^b}{c^2}}\left(n - {\frac 1 n}\right)\right],\\
{\stackrel {o}{\gamma}}{}^a{}_b &=& {\frac \lambda {1 - v^2/c^2}}\,\left(
n - {\frac 1 n}\right)\,{}^{(3)}\eta^a{}_{cb}\,{\frac {v^c}{c}}.
\end{eqnarray}
Here $\lambda := \sqrt{\varepsilon\varepsilon_0/\mu\mu_0}$ and $n := 
\sqrt{\varepsilon\mu}$ is the refraction index of the medium. The 3-dimensional
indices are raised and lowered with the help of the 3-dimensional spacetime
metric ${}^{(3)}g_{ab}$, see the line element (\ref{metM}). 

The magnetoelectric contributions to the constitutive matrices read
\begin{eqnarray}
{\stackrel {\rm me}{\varepsilon}}{}^{ab} &=& -\,{\frac {2}{1 - v^2/c^2}}
\,\underline{\beta}{}^{(a}{}_c\,\epsilon^{b)cd}\,v_d/c^2,\\
({\stackrel {\rm me}{\mu}}{}^{-1})_{ab} &=& -\,{\frac {2}{1 - v^2/c^2}}
\,\underline{\beta}{}^c{}_{(a}\,\epsilon_{b)cd}\,v^d,\\
{\stackrel {\rm me}{\gamma}}{}^a{}_b &=&  {\frac {1}{1 - v^2/c^2}}\,
\widehat{\beta}{}^a{}_b.\label{game} 
\end{eqnarray}
Here we introduced the notation
\begin{equation}
\widehat{\beta}{}^a{}_b :=  \underline{\beta}{}^a{}_b - {\frac 1{c^2}}
\,\epsilon^{acd}\,\epsilon_{bc'd'}\,\underline{\beta}{}^{c'}{}_c\,v^{d'}v_d,
\end{equation}
with the ``doubly projected" magnetoelectric matrix
\begin{equation}
\underline{\beta}{}^a{}_b := \left(\delta^a_c - {\frac 1{c^2}}\,v^av_c\right)
\left(\delta^d_b - {\frac 1{c^2}}\,v^dv_b\right)\widetilde{\beta}^c{}_d.
\end{equation}

For the medium at rest (with $v^a = 0$), we straightforwardly verify that
the above formulas reduce to the constitutive law
\begin{eqnarray}
{\cal D}^a &=& \varepsilon\varepsilon_0\,E^a + \beta^a{}_b\,B^b,\\
{\cal H}_a &=& -\,\beta^b{}_a\,E_b + {\frac {1}{\mu\mu_0}}\,B_a. 
\end{eqnarray}
{}From these, the constitutive relations of the moving magnetoelectric medium
(\ref{Atot})-(\ref{game}) can be alternatively derived with the help of the
transformation (\ref{dsigmaM}) that links the laboratory to the material 
foliations. 

The new magnetoelectric term in the Lagrangian (\ref{genLem}) will modify the 
equations of motion of the medium and the total energy-momentum of the system. 
Direct computation yields:
\begin{eqnarray}
T_i{}^j &=& -\,\delta_i^j\,p_{\rm eff} + {\frac {p_{\rm eff} 
+ \rho_{\rm eff}}{c^2}}\,u_iu^j + {\frac 2c}\,q_{(i}u^{j)}\nonumber\\
&& + {\frac 1 {\mu_0\mu}}\left[-\,F_{ik}\,F^{jk} + {\frac 1 4}\,\delta_i^j
\,F_{kl}F^{kl} + {\frac {1 - n^2}{c^2}}\,F_{ik}u^kF^{jl}u_l\right].\label{genTeff}
\end{eqnarray}
Here the effective energy density and the effective pressure remain the same 
(\ref{Peff}), (\ref{Reff}), whereas the vector $q_i$ is defined by
\begin{equation}\label{q}
q_i = {\frac 12}\,\beta^{i'}{}_{j'}\,P^j_i\eta^{j'}{}_j{}^{lm}F_{lm}F_{i'k}u^k.
\end{equation}
Clearly, it is orthogonal to the velocity of the medium, $u^iq_i = 0$. The
terms with the structure $2u_{(i}q_{j)}/c$ are well known in the models of
relativistic fluids; they usually describe fluids with fluxes of energy. It
is interesting that the magnetoelectric parameters introduce such an 
{\it additional energy flux} term in the total energy-momentum tensor. 

At the beginning, we assumed that the axion is absent. However, it is instructive
to check how it could contribute to the energy-momentum. For this, we relax for
a moment the tracefree condition for the magnetoelectric tensor and allow for
a nontrivial trace $\beta^i{}_j = \delta^i_j\,\widetilde{\alpha}$. This yields
a contribution $\sim {\frac 12}\,\widetilde{\alpha}\,P^j_i\eta^{j'}{}_j{}^{lm}
F_{lm}F_{j'k}u^k$ in (\ref{q}). With the help of the identity $2u^{[k}
\eta^{j']}{}_j{}^{lm} \equiv u_j\eta^{j'klm} + u^l\eta^{j'}{}_j{}^{km} +
u^m\eta^{j'}{}_j{}^{lk}$, this is transformed into ${\frac 12}\,\widetilde{\alpha}
\,P^j_iu_j\eta^{j'klm}F_{lm}F_{j'k}$. It vanishes because of the identity
$P^j_iu_j\equiv 0$. In other worlds, this demonstrates that the {\it axion 
does not contribute to the energy-momentum}, in accordance with its general 
theory \cite{HO02}.

\subsection{Explicit components of the energy-momentum}

All the results of the previous two subsections 
are {\it generally covariant}, i.e., they are valid in an arbitrary curved
spacetime with any metric $g_{ij}$. Now, returning to a more narrow case
actually discussed by Feigel, we will specialize to the Minkowski spacetime
with $g_{ij} = {\rm diag}(c^2, -1, -1, -1)$. 

In order to write down the separate components of the energy-momentum
tensor, we first notice that $u^0 = \gamma, u^a = \gamma\,v^a$, with the 
3-velocity $v^a$ and the usual relativistic (Lorentz) factor $\gamma = 1/\sqrt{1 
- v^2/c^2}$. Then, denoting $X_i := F_{ij}u^j$, we find 
\begin{equation}
X_0 = -\gamma (vE),\qquad X_a = \gamma ({\bf E} + {\bf v}\times{\bf B})_a.
\end{equation}
Accordingly, we have the invariant [that enters (\ref{Peff}), (\ref{Reff})]
\begin{equation}
X_iX^i = - \gamma^2\left[E^2 - (vE)^2/c^2 + v^2B^2 - (vB)^2 + 2({\bf E}\cdot
[{\bf v}\times{\bf B}])\right].\label{X2}
\end{equation}
Here we denoted the scalar product $(vE) = ({\bf v}\cdot{\bf E})$ and
similarly $(vB)$. 

After these preliminaries, we find, componentwise, the {\it energy density}
\begin{eqnarray}
T_0{}^0 = u &=& \gamma^2\left[\rho_{\rm eff} + {\frac {v^2}{c^2}}p_{\rm eff}
+ {\frac {\varepsilon_0(1 - n^2)}{\mu c^2}}(vE)^2\right]\nonumber\\ 
&&\qquad +\,{\frac 1 {2\mu}}\left(\varepsilon_0\,E^2 + {\frac 1 {\mu_0}}
\,B^2\right) - {\frac {2\gamma v^a}{c}}\,q_a,
\end{eqnarray}
the {\it energy flux density} (or Poynting vector)
\begin{eqnarray}
T_0{}^a = s^a &=& \gamma^2\left[(\rho_{\rm eff} + p_{\rm eff})\,{\bf v} 
+ {\frac {\varepsilon_0(1 - n^2)}\mu}(vE)({\bf E} + {\bf v}\times{\bf B})
\right]^a\nonumber\\ &&\qquad\qquad +\,{\frac 1 {\mu\mu_0}}\,\left[{\bf E}
\times{\bf B}\right]^a - \gamma c\left(\delta^a_b - {\frac 1{c^2}}v^av_b\right)
q^b,\label{sa}
\end{eqnarray}
the {\it momentum density}
\begin{eqnarray}
T_a{}^0 = -\,p_a &=& -\,\gamma^2\left[{\frac {\rho_{\rm eff} + p_{\rm eff}}
{c^2}}\,{\bf v} + {\frac {\varepsilon_0(1 - n^2)}{\mu c^2}}(vE)({\bf E} + 
{\bf v}\times{\bf B})\right]_a\nonumber\\ &&\qquad\qquad 
+\,{\frac {\varepsilon_0}\mu}\,\left[{\bf B}\times{\bf E}\right]_a 
+ {\frac {\gamma}{c}}\left(\delta^b_a - {\frac 1{c^2}}v^bv_a\right)q_b,\label{pa}
\end{eqnarray}
and the {\it stress} tensor
\begin{eqnarray}
T_a{}^b = S_a{}^b &=& -\,p_{\rm eff}\left(\delta_a^b + {\frac {\gamma^2}{c^2}}
\,v_av^b\right) - \gamma^2{\frac {\rho_{\rm eff}}{c^2}}\,v_av^b\nonumber\\
&& +\,{\frac {\varepsilon_0}\mu}\left(E_aE^b - {\frac 1 2}\,\delta_a^b\,E^2
\right) + {\frac 1 {\mu\mu_0}}\left(B_aB^b - {\frac 1 2}\,\delta_a^b\,B^2
\right)\nonumber\\
&& -\,{\frac {\varepsilon_0(1 - n^2)}\mu}\,\gamma^2\,({\bf E} + {\bf v}\times
{\bf B})_a({\bf E} + {\bf v}\times{\bf B})^b - 2\gamma q_{(a}v^{b)}.\label{Sab}
\end{eqnarray}
As we see, the vector $q_a$ indeed induces an additional energy flux in
(\ref{sa}) and adds a corresponding magnetoelectric term in the electromagnetic
momentum density (\ref{pa}).

For completeness, let us also give the explicit expressions for the 
{\it effective} energy density and the {\it effective} pressure of the medium: 
\begin{eqnarray}
\rho_{\rm eff} &=& \rho - {\frac {\varepsilon_0(1 - n^2)}{2\mu}}\gamma^2\Big[
E^2 - (vE)^2/c^2 + v^2B^2 - (vB)^2 +\,2({\bf E}\cdot[{\bf v}\times{\bf B}])\Big],\\
p_{\rm eff} &=& p - {\frac {\varepsilon_0(1 - n^2)}{2\mu}}\gamma^2\Big[
E^2 - (vE)^2/c^2 + v^2B^2 - (vB)^2 +\,2({\bf E}\cdot[{\bf v}\times{\bf B}])\Big].
\end{eqnarray}

The energy-momentum tensor obtained here is different from both the Minkowski
and the Abraham energy-momentum tensors. It is thus necessary to check whether
the new expression is consistent with the main experiments in phenomenological 
electrodynamics. As a first step, we consider usual matter without 
magnetoelectric properties.

\section{Experiments with light pressure}\label{Exp}

In order to test how the formalism works, we analyse in this section the famous
experiments of Richards and Jones \cite{Jones1,Jones2}. Their measurements have
shown that the radiation pressure on a metallic plate immersed into a dielectric 
fluid is proportional to the refraction index $n$ of the medium. 

The problem of the propagation of a plane wave in a dielectric medium is solved
exactly. Suppose a wave travels along the $x$ axis, with the metal plate 
surface located at $x =0$. For $x < 0$, we have dielectric fluid with the 
permeability and permittivity $\varepsilon_1, \mu_1$, and for $x > 0$ we have 
the metal characterized by $\varepsilon_2, \mu_2$ and by the electric conductivity 
$\sigma$. Then the electric and magnetic fields that describe the plane waves
normally incident on and reflected form the metal surface are ($x \leq 0$)
\begin{eqnarray}
E_y &=& a\,e^{-i\omega t}\left(e^{i{\frac {\omega n_1}c}x} +
{\cal R}\,e^{-\,i{\frac {\omega n_1}c}x}\right),\\ 
B_z &=& {\frac {n_1} c}\,a\,e^{-i\omega t}\left(
e^{i{\frac {\omega n_1}c}x} - {\cal R}\,e^{-\,i{\frac {\omega n_1}c}x}\right).
\end{eqnarray}
The fields in the metal ($x\geq 0$) read
\begin{eqnarray}
E_y &=& {\cal T}\,a\,e^{-i\omega t}
\,e^{i{\frac {\omega n_2}c}{\cal K}x},\\ 
B_z &=& {\frac {n_2} c}\,{\cal T}\,{\cal K}\,a\,e^{-i\omega t}
\,e^{i{\frac {\omega n_2}c}{\cal K}x}.
\end{eqnarray}
Without loosing generality, we assume that the electric field is directed
along the $y$ axis. 
We use the complex representation to simplify the formulas. Here the complex
variable $a$ describes the amplitude of the incident wave, whereas the complex
reflection and transmission coefficients are denoted ${\cal R}$ and ${\cal T}$,
respectively. As usual, $n_1 = \sqrt{\varepsilon_1\mu_1}$ and $n_2 = 
\sqrt{\varepsilon_2\mu_2}$. Solving the Maxwell equations, we find 
${\cal K} = k + i\delta$ with 
\begin{equation}
k = \left[{\frac 1 2}\left(\sqrt{1 + \left({\frac \sigma {\varepsilon
\varepsilon_0\,\omega}}\right)^2} \, + 1\right)\right]^{1/2},\quad
\delta = \left[{\frac 1 2}\left(\sqrt{1 + \left({\frac \sigma {\varepsilon
\varepsilon_0\,\omega}}\right)^2} \, - 1\right)\right]^{1/2}.
\end{equation}
{}From the matching conditions at the boundary $x = 0$, we obtain explicitly
\begin{equation}
{\cal R} = {\frac {\sqrt{\frac {\varepsilon_1}{\mu_1}} - {\cal K}
\,\sqrt{\frac {\varepsilon_2}{\mu_2}}}{\sqrt{\frac {\varepsilon_1}{\mu_1}} 
+ {\cal K}\,\sqrt{\frac {\varepsilon_2}{\mu_2}}}},\qquad {\cal T} = 
{\frac {2\sqrt{\frac {\varepsilon_1}{\mu_1}}} {\sqrt{\frac {\varepsilon_1}{\mu_1}} 
+ {\cal K}\,\sqrt{\frac {\varepsilon_2}{\mu_2}}}}\,.
\end{equation}
All the formulas are simplified in the case of the strongly conducting matter,
when $\sigma \gg \varepsilon\varepsilon_0\,\omega$ (see \cite{Brevik}, e.g.),
but this condition is actually not necessary in our analysis. 

It is now straightforward to compute the force per area of the metal surface,
which is most conveniently given in terms of the stress tensor components,
$f_x = -\,S_x{}^x$. We put $v^a = 0$ since the dielectric fluid is at rest. 
Then (\ref{Sab}) yields
\begin{eqnarray}
f_x &=& p + {\frac {\varepsilon_1\varepsilon_0}{2}}\,|E|^2 
+ {\frac 1{2\mu_1\mu_0}}\,|B|^2\nonumber\\
&=& p + {\frac {n_1}{c}}\,(1 + |{\cal R}|^2)\,s_{(i)}^x.\label{fx}
\end{eqnarray}
Here the right-hand side is evaluated at $x=0$, and we used (\ref{sa}) to 
substitute the energy flux (Poynting vector) of the incident wave $s_{(i)}^x$
along the $x$ axis. As usual, the averaging over a period was performed in 
these calculations. The hydrodynamical pressure term $p$ is obviously cancelled
by the same fluid pressure force acting on the metal plate from the opposite 
side. So, the net force per area on the is described only by the last term 
in (\ref{fx}), in complete agreement with the experimental results (cf. also 
with the formula (4.22) in \cite{Brevik}). 

Another application of our formalism to the analysis of the important 
experiments of Ashkin and Dziedzic \cite{Ashkin1,Ashkin2} will be considered
elsewhere.

\section{Discussion and conclusion}\label{Concl}

In this paper we have developed a model for a moving electrodynamical medium
within the framework of a variational approach. The application to the 
observations of Richards and Jones \cite{Jones1,Jones2} demonstrates the 
viability of the model. Our next aim is to use this model for the analysis
of the possibility of the Feigel effect  \cite{feigel} predicted recently.

We have derived the total energy-momentum of moving matter and the 
electromagnetic field. In particular, we found that the magnetoelectric 
properties contribute with the vector $q_i$ to an additional energy flux
and momentum density. 

We are now in a position to compare the quantities derived 
here with those used by Feigel \cite{feigel}. A direct inspection shows that 
the expression of Feigel is not obtained from the formulas above in the 
non-relativistic limit, when we assume that $v^2/c^2 \ll 1$. This fact casts 
certain doubts on the validity of the original conclusions of Feigel. 

In the meantime, an interesting new paper \cite{Birkenland} appeared where
the Feigel effect was analysed with the help of a Green-function techniques.
A similar discussion, which is based on the quantum-mechanical formalism and 
using the above variational formalism, will be published elsewhere.

\end{document}